\documentclass[10pt,twoside,twocolumn,american,eqsecnum]{revtex4}
\usepackage[T1]{fontenc}
\usepackage[latin1]{inputenc}
\usepackage{amsmath}
\usepackage{graphicx}

\makeatletter
\@addtoreset{equation}{section} 

\usepackage{amssymb}

\usepackage{babel}
\makeatother
\begin{document}

\title{Can dark energy evolve to the Phantom? }

\author{Alexander Vikman}

\address{Ludwig-Maximilians-Universität München, Department of Physics, Theresienstraße
37, D-80333 Munich, Germany }

\begin{abstract}
Dark energy with the equation of state $w(z)$ rapidly evolving from
the dustlike ($w\simeq0$ at $z\sim1$) to the phantomlike ($-1.2\lesssim w\lesssim-1$
at $z\simeq0$) has been recently proposed as the best fit for the
supernovae Ia data. Assuming that a dark energy component with an
arbitrary scalar-field Lagrangian $p(\varphi,\nabla_{\mu}\varphi)$
dominates in the flat Friedmann universe, we analyze the possibility
of a dynamical transition from the states $(\varphi,\dot{\varphi})$
with $w\geq-1$ to those with $w<-1$ or vice versa. We have found
that generally such transitions are physically implausible because
they are either realized by a discrete set of trajectories in the
phase space or are unstable with respect to the cosmological perturbations.
This conclusion is confirmed by a comparison of the analytic results
with numerical solutions obtained for simple models. Without the assumption
of the dark energy domination, this result still holds for a certain
class of dark energy Lagrangians, in particular, for Lagrangians quadratic
in $\nabla_{\mu}\varphi$. The result is insensitive to topology of
the Friedmann universe as well.
\end{abstract}
\maketitle

\section{Introduction}

One of the greatest challenges in modern cosmology is understanding
the nature of the observed late-time acceleration of the universe.
The present acceleration expansion seems to be an experimental fact,
now that data from supernovae type Ia \cite{Perlmutter,Riess}, corroborated
later by those from the cosmic microwave background \cite{WMAP},
have been recently confirmed by the observations of the largest relaxed
galaxy clusters \cite{Chandra}. Although the observations are in
good agreement with the simplest explanation given by a cosmological
constant $\Lambda$ of order $(10^{-3}\,\textrm{eV})^{4}$, the mysterious
origin of this tiny number, which is about 120 orders smaller than
the naive expectations, gives rise to the idea of a dynamical nature
of this energy. Possible dynamical explanations of this phenomenon
are given in various frameworks. One of them is known as quintessence
(see e.g. \cite{qiuntessence} and other references from the review
\cite{Review}). In this framework the equation of state $p=w\varepsilon$
is such that $w\geq-1$. Another proposal is the phantom scalar fields
(see e.g.~\cite{Caldwell Manace}) which possess the super-negative
equation of state $w\leq-1$, due to the {}``wrong'' sign before
the kinetic term in the Lagrangian. Alternatively, there is a more
general possibility under the name $k-$essence \cite{k essence first,k-Essence,Mukhanov}
which is an effective scalar-field theory described by a Lagrangian
with a nonlinear kinetic term. For this model, the equation of state
$w$ is not constrained to be larger or smaller than $-1$. Allowing
the dark energy to be dynamical provides an opportunity to study the
so-called coincidence problem which asks why dark energy domination
begins just at the epoch when sentient beings are able to observe
it. The main advantage of $k-$essence is its ability to solve this
problem in a generic way (for details see \cite{k-Essence}), whereas
the first two models require a fine-tuning of parameters. 

Without imposing the prior constraint $w\geq-1$, the observations
seem to favor the dark energy with the present equation of state parameter
$w<-1$ (see e.g.~Ref.~\cite{Chandra,Riess New,Mersini,Sami1}).
Moreover, recently it was argued (see Ref.~\cite{Starobinsky,Starobinsky ujjani}
and other constraints on $w(z)$ obtained in Refs.~\cite{Corasaniti,Hannestad,Padmanabhan,Padnamadhan2,Xinmin Zhang,Huterer})
that the dark energy with the equation of state parameter $w(z)$
rapidly evolving from the dustlike $w\simeq0$ at high redshift $z\sim1$,
to phantomlike $-1.2\lesssim w\lesssim-1$ at present $z\simeq0$,
provides the best fit for the supernovae Ia data and their combinations
with other currently available data from the measurements of cosmic
microwave background radiation (CMBR) and from 2dF Galaxy Redshift
Survey (2dFGRS). 

Matter with $w<-1$ violates the dominant energy condition which is
a sufficient condition of the conservation theorem \cite{Hawking}.
Therefore for such models one cannot guarantee the stability of vacuum
on the classical level. The instability can reveal itself at the quantum
level as well. In fact, it was shown that the phantom scalar fields
are quantum-mechanically unstable with respect to decay of the vacuum
into gravitons and phantom particles with negative energy \cite{Hoffman,Cline}.
Assuming that the phantom dark energy is an effective theory allows
one to escape this problem through the appropriate fine-tuning of
a cutoff parameter. If the dark energy could dynamically change its
equation of state from a phantomlike one to that with $w\geq-1$,
then this transition would prevent the undesirable particle production
without such a fine-tuning. Here it is worth mentioning that quantum
effects on a locally de Sitter background could lead to the effective
parameter $w<-1$ (see Ref.~\cite{Onemli,Onemli 2}). 

Another fundamental physical issue where this transition could play
an important role is the cosmological singularity problem. If $w<-1$
in an expanding Friedmann universe, then the positive energy density
of such phantom matter generally becomes infinite in finite time,
overcoming all other forms of matter and, hence, leads to the late-time
singularity called the {}``big rip'' \cite{Big rip}. The transition
under consideration could naturally prevent this late-time singularity.
Here it is worthwhile to mention that for certain potentials and initial
conditions the phantom scalar fields can escape this singularity by
evolving to a late-time asymptotic which is the de Sitter solution
with $w=-1$ \cite{Sami 2,Sami1}. Moreover, it was argued that the
quantum effects can prevent the developing of the {}``big-rip''
singularity as well \cite{Quant Escape}. 

On the other hand, to avoid the big crunch singularity, which arises
in various pre-big bang and cyclic scenarios (see e.g.~\cite{Cyclic,Tolman,PRe big bang}),
one assumes that the universe can bounce instead of collapsing to
the singularity. The existence of a nonsingular bouncing solution
in a flat (or open) Friedmann universe ($k\neq+1$) requires the violation
of the null energy condition ($\varepsilon+p\geq0$) during the bounce
\cite{Bounce}. If the energy density $\varepsilon$ is constrained
to be positive, then it follows that $w<-1$ is the necessary condition
for the bounce. But the energy density of such phantom matter would
rapidly decrease during the collapse and therefore only the transition
from $w\geq-1$ to $w<-1$ just before the bounce could explain the
nonsingular bouncing without a fine-tuning in initial energy densities
of phantom and other forms of matter present in the universe. 

It is worth noting as well that for regimes where the equation of
state of the $k-$essence field is greater than $-1$ it is possible
to find a quintessence model which gives the same cosmological evolution
but behaves differently with respect to cosmological perturbations
\cite{Liddle}. Hence, it is interesting whether this equivalence
can be broken dynamically.\\

In this paper we consider the cosmological dynamics of a $k-$essence
field $\varphi$, described by a general Lagrangian $p$ which is
a local function of $\varphi$ and $\nabla_{\mu}\varphi$. The Lagrangian
depends only on $\varphi$ and a scalar quantity, \begin{equation}
X\equiv\frac{1}{2}\nabla_{\mu}\varphi\nabla^{\mu}\varphi.\end{equation}
 First of all, we determine the properties of a general Lagrangian
$p(\varphi,X)$, which are necessary for the smooth transition of
the dark energy from the equation of state $w(\varphi,X)\geq-1$ to
$w(\varphi,X)<-1$ or vice versa. The transition obviously happens
if the system passes through the boundaries of the domains in the
space $(\varphi,X)$, defined by these inequalities. In most of the
paper, we assume that the dark energy dominates in a spatially flat
Friedmann universe. The main question is whether trajectories connecting
these domains on the phase space $(\varphi,\dot{\varphi})$ exist
and are stable with respect to cosmological perturbations. In the
case of the phase curves which do not violate the stability conditions,
we study their asymptotic behavior in the neighborhood of the points
where the transition could occur. To proceed with this analysis, we
linearize the equation of motion in the neighborhood of these points
and then use the results of the qualitative theory of differential
equations. For the dark energy models described by Lagrangians linear
in $X$, we perform this investigation beyond the linear approximation.
For this class of Lagrangians, we illustrate the outcome of our analysis
by numerically obtained phase curves. Finally, we generalize the results
to the cases of spatially not-flat Friedmann universes filled with
a mixture or the dark energy and other forms of matter.

\section{General framework }

Assuming the dominance of the dark energy, we neglect all other forms
of matter and consider a single scalar field $\varphi$ interacting
with gravity. After all, we will see that the results can be easily
extended to the models with additional forms of matter. The action
of the model reads in our units ($M_{p}=\hbar=c=1$, where $M_{p}$
is the reduced Planck mass~$M_{p}=(8\pi G)^{-1/2}=1.72\times10^{18}$GeV
) as follows:\begin{equation}
S=S_{g}+S_{\varphi}=\int d^{4}x\sqrt{-g}\left[-\frac{R}{2}+p(\varphi,X)\right],\end{equation}
where $R$ is the Ricci scalar and $p(\varphi,X)$ is the Lagrangian
density for the scalar field. This kind of action may describe a fundamental
scalar field or be a low-energy effective action. In principle, the
Lagrangian density $p(\varphi,X)$~can be non-linear on $X$. For
example, in string and supergravity theories nonlinear kinetic terms
appear generically in the effective action describing moduli and massless
degrees of freedom due to higher order gravitational corrections to
the Einstein-Hilbert action \cite{Polchinsky,Gross}. The {}``matter''
energy-momentum tensor reads\begin{eqnarray}
T_{\mu\nu}\equiv\frac{2}{\sqrt{-g}}\left[\frac{\delta S_{\varphi}}{\delta g^{\mu\nu}}\right]\label{eq:Energy-Momentum}\\
=p_{,X}(\varphi,\, X)\nabla_{\mu}\varphi\nabla_{\nu}\varphi-p(\varphi,\, X)g_{\mu\nu}.\nonumber \end{eqnarray}
Here a comma denotes a partial derivative with respect to $X$. The
last equation shows that, if $\nabla_{\nu}\varphi$ is timelike (i.e.
$X>0$), the energy-momentum tensor is equivalent to that of a perfect
fluid, \begin{equation}
T_{\mu\nu}=(\varepsilon+p)U_{\mu}U_{\nu}-pg_{\mu\nu},\label{eq:energy momentum tensor}\end{equation}
 with pressure $p(\varphi,X)$, energy density \begin{equation}
\varepsilon(\varphi,X)=2Xp_{,X}(\varphi,X)-p(\varphi,X),\label{eq:energy density on X}\end{equation}
and four velocity \begin{equation}
U_{\mu}=\frac{\nabla_{\mu}\varphi}{\sqrt{2X}}.\end{equation}
The equation of motion for the scalar field can be obtained either
as a consequence of the energy-momentum tensor conservation $\nabla_{\mu}T_{\nu}^{\mu}=0$
or directly from the extremal principle $\delta S_{\varphi}/\delta\varphi=0$:\begin{equation}
p_{,X}\square_{g}\varphi+p_{,XX}\left(\nabla_{\mu}\nabla_{\nu}\varphi\right)\nabla^{\mu}\varphi\nabla^{\nu}\varphi+\varepsilon_{,\varphi}=0,\label{eq:general equation of motion}\end{equation}
where $\square_{g}\equiv g_{\mu\nu}\nabla^{\mu}\nabla^{\nu}$ and
$\nabla^{\mu}$ denotes the covariant derivative. For this fluid we
can define the equation of state parameter $w$ as usual:\begin{equation}
w\equiv\frac{p}{\varepsilon}.\label{eq:W for P}\end{equation}
There is increasing evidence that the total energy density of the
universe is equal to the critical value, and hence in the most part
of the paper we will consider a flat Friedmann universe. In the end,
we shall show that the results are also applicable in the cases of
closed and open universes. Thus, the background line element reads\begin{equation}
ds^{2}=g_{\mu\nu}dx^{\mu}dx^{\nu}=dt^{2}-a^{2}(t)\, d\mathbf{x}^{2}.\label{eq:4d line element}\end{equation}
The Einstein equations can be written for our background in the familiar
form: \begin{equation}
\frac{\ddot{a}}{a}=-\frac{1}{6}\left(\varepsilon+3p\right),\label{Einstein's equation 1}\end{equation}
\begin{equation}
H^{2}=\frac{\varepsilon}{3},\label{Einstein's equation 2}\end{equation}
where $H\equiv\dot{a}/a$ is the Hubble parameter and a dot denotes
derivative with respect to the physical time $t$. These equations
also imply a continuity equation:\begin{equation}
\dot{\varepsilon}=-3H(\varepsilon+p).\label{eq:continuity}\end{equation}
 In general, whenever $\dot{a}\neq0$ any two of these three last
equations imply the third one (by compatible initial conditions).
Usually it is easier to work with the second and the third equations
(these are the Friedmann equations). Note that, from Eq.~(\ref{Einstein's equation 2}),
$\varepsilon$ was constrained to be non-negative. 

Because of the homogeneity and isotropy of the background, we get
$X=\frac{1}{2}\dot{\varphi}^{2}$ and $p_{,\dot{\varphi}}=\dot{\varphi}p_{,X}$
so the energy density looks as the energy in usual 1D classical mechanics\begin{equation}
\varepsilon(\varphi,\dot{\varphi})=\dot{\varphi}p_{,\dot{\varphi}}-p.\label{eq:enrgy density on t}\end{equation}
 Expressing $H$ from the first Friedmann equation (\ref{Einstein's equation 2}),
we can rewrite Eq.~(\ref{eq:general equation of motion}) in the
case of the homogeneous and isotropic flat background as follows:\begin{equation}
\ddot{\varphi}\varepsilon_{,X}+\dot{\varphi}p_{,X}\sqrt{3\varepsilon}+\varepsilon_{,\varphi}=0.\label{eq:Eq. of moution}\end{equation}
So far as $\dot{a}(t)\neq0$, all of the information about the dynamics
of gravity and scalar field is contained in the equation written above.
In accordance with our initial simplification the dark energy should
dominate in the universe; therefore we assume throughout the paper
that $\varepsilon>0$. 

Following \cite{Perturbations}, we introduce the effective sound
speed of the perturbations, \begin{equation}
c_{s}^{2}\equiv\frac{p_{,X}}{\varepsilon_{,X}}.\label{eq:saund speed}\end{equation}
Then the equation of motion takes the form\begin{equation}
\ddot{\varphi}+\dot{\varphi}c_{s}^{2}\sqrt{3\varepsilon}+\frac{\varepsilon_{,\varphi}}{\varepsilon_{,X}}=0.\end{equation}
In most of this paper, we shall assume that the solutions $\varphi(t)$
and Lagrangians $p(\varphi,X)$ have enough continuous derivatives.
So, for example, $\varphi(t)$ will be mostly considered as being
at least of the class $C^{2}:$~$\varphi(t)$, $\dot{\varphi}(t)$,
$\ddot{\varphi}(t)$ are continuous.

\section{Possible Mechanisms of the transition}

There are two possibilities for the evolution of dark energy from
$w\geq-1$ to a phantom dark energy with $w<-1$ (or vise versa).
These are a continuous transition, in which the dark energy evolves
through points where $w=-1$, and a discontinuous transition occurring
through points where $\varepsilon=0$, provided that the pressure
$p$ is finite. Since by assumption the dark energy is the dominating
source of gravitation, we cannot have $\varepsilon=0$ and therefore
it is sufficient to consider only continuous transitions.\\

Further, throughout the paper we will usually suppose that for the
dynamical models under consideration there exist solutions $\varphi(t)$
and corresponding to them moments of time $t_{c}$ such that \begin{equation}
w\left[\varphi(t_{c}),X(t_{c})\right]=-1.\end{equation}
Henceforth the index $c$ denotes a physical quantity taken at $t_{c}$;
i.e., $\varphi_{c}\equiv\varphi(t_{c})$, $\varepsilon_{c}\equiv\varepsilon(t_{c})$,
etc. \\

The parameter $w$ can be expressed with the help of Eq.~(\ref{eq:energy density on X})
in the following form, more convenient for a study of continuous transitions:\begin{equation}
w=-1+\frac{2X}{\varepsilon}p_{,X}.\label{eq:W on fi dot}\end{equation}
 Since $X=\frac{1}{2}\dot{\varphi}^{2}\geq0$ and $\varepsilon>0$,
we find that $w<-1$ corresponds to $p_{,X}<0$, whereas $w>-1$ implies
$p_{,X}>0$. In accordance with our notation, the equation of state
parameter $w$ takes the value $-1$ at the points $\Psi_{c}\equiv\left(\varphi_{c},\dot{\varphi}_{c}\right)$,
where either $X=0$ or $p_{,X}=0$. Because of these equations, the
points $\Psi_{c}$ generally form curves $\gamma(\lambda)$ and isolated
points in the phase space $(\varphi,\dot{\varphi})$ of the dynamical
system given by Eq.~(\ref{eq:Eq. of moution}). The curves $\gamma(\lambda)$
may intersect.\\

For our purposes, it is convenient to divide the set of points $\Psi_{c}$
into three disjunct subsets: 

\begin{description}
\item [A)]The $\varphi-$axis of the phase plot $(\varphi,\dot{\varphi})$,
i.e., $\dot{\varphi}_{c}=0$.
\item [B)]The points where $p_{,X}(\Psi_{c})=0$ but $\dot{\varphi}_{c}\neq0$
and\\
$\varepsilon_{,X}(\Psi_{c})\neq0$.
\item [C)]The points where $p_{,X}(\Psi_{c})=0$ and $\varepsilon_{,X}(\Psi_{c})=0$
but $\dot{\varphi}_{c}\neq0$.
\end{description}
Further in this section we will study the dynamics of the scalar field
$\varphi$ in the neighborhoods of $\Psi_{c}$ separately for these
cases. If the system evolves from the states $(\varphi,\dot{\varphi})$
where $w\geq-1$ to the states with $w<-1$ (or vise versa), the function
$p_{,X}$ changes sign.\\

It is worth noting that, if the scalar dark energy were equivalent
to an {}``isentropic'' fluid for which the pressure $p$ is a function
only of $\varepsilon$, then the possibility of evolving through the
points $\varepsilon_{c}$ where $w(\varepsilon_{c})=-1$ could be
easily ruled out. Indeed, in that case we could rewrite the continuity
equation (\ref{eq:continuity}) only in terms of $\varepsilon$:\begin{equation}
\dot{\varepsilon}=-\sqrt{3\varepsilon}\left[\varepsilon+p(\varepsilon)\right],\label{eq:muster}\end{equation}
so that the system of Einstein equations (\ref{Einstein's equation 1})
and (\ref{Einstein's equation 2}) could be reduced to Eq.~(\ref{eq:muster})
and the values of energy density $\varepsilon_{c}$ would be fixed
points of this equation. A dynamical transition through a fixed point
is clearly impossible. 

An example of the dark energy which seems to be equivalent to the
{}``isentropic'' fluid is the simple model described by the Lagrangian
$p=p(X)$ depending only on $X$. Let us further assume that there
are some values $X_{c}$, where $w(X_{c})=-1$. If Eq.~(\ref{eq:energy density on X})
is solvable with respect to $X$ in the neighborhoods of these points
$X_{c}$, then one can find $X(\varepsilon)$ and therefore the pressure
is a function only of energy density $p(\varepsilon)\equiv p\left[X(\varepsilon)\right]$.
Thus the system is equivalent to the {}``isentropic'' fluid, $X_{c}$
are fixed points, and the transition through $w=-1$ is impossible.
It remains to consider the conditions on the function $\varepsilon(X)$
under which Eq.~(\ref{eq:energy density on X}) is solvable with
respect to $X$. From the theorem about the inverse function, Eq.~(\ref{eq:energy density on X})
is solvable with respect to $X$ if \begin{equation}
\varepsilon_{,X}(X_{c})=\left[2Xp_{,XX}(X_{c})+p_{,X}(X_{c})\right]\neq0.\label{eq:cond p(x)}\end{equation}
One can see directly from the equation of motion (\ref{eq:Eq. of moution})
and condition (\ref{eq:cond p(x)}) that $X_{c}$ are fixed point
solutions. In fact, as it was shown in Ref.~\cite{scherrer}, there
generally exists the solution $X(t)\equiv X_{c}$ and moreover it
is an attractor in an expanding Friedmann universe. Thus, the transition
is generally forbidden for systems described by purely kinetic Lagrangians
$p(X)$. 

In the general case when $p=p(\varphi,X)$, the pressure cannot be
expressed only in terms of $\varepsilon,$ since $\varphi$ and $X$
are independent.

\subsection{\label{sub:Tansition-at-X=3D0}Transition at points $X_{c}=0$}

Here we will analyze the possibility of the transition in the case
(A) $\dot{\varphi}_{c}=0$. Namely, we are going to study the properties
of the solutions $\varphi(t)$ in the neighborhood of the line $\dot{\varphi}_{c}=0$.
Differentiating the equation of state parameter with respect to the
time, we have \begin{equation}
\dot{w}=\frac{2\dot{X}}{\varepsilon}p_{,X}+\frac{2X}{\varepsilon}\dot{p_{,X}}-\frac{2X}{\varepsilon^{2}}p_{,X}\dot{\varepsilon}.\label{eq:wdot}\end{equation}
At the points under consideration we have $\dot{w}_{c}=0$ because
$X_{c}=0$ and, respectively, $\dot{X}_{c}=\dot{\varphi}_{c}\ddot{\varphi}_{c}=0$.
Moreover, the time derivatives in the second and third summands vanish
at these points as well due to the continuity equation (\ref{eq:continuity})
and the formula\begin{equation}
\dot{p_{,X}}=\dot{\varphi}\left(p_{,\varphi X}+\ddot{\varphi}p_{,XX}\right).\label{eq:dpx/dt}\end{equation}
 Let us differentiate the $\dot{w}$ once more with respect to the
time. The only term which survives from the formula (\ref{eq:wdot})
at the points $X_{c}=0$ is the first term. Hence, we have\begin{equation}
\ddot{w}_{c}=\left[\frac{2\ddot{X}}{\varepsilon}p_{,X}\right]_{c}=\left[\frac{2\ddot{\varphi}^{2}}{\varepsilon}p_{,X}\right]_{c}.\label{eq:w two dot}\end{equation}
 Using the equation of motion (\ref{eq:Eq. of moution}), we can express
$\ddot{\varphi}$ through the $p$ and its derivatives \begin{equation}
\ddot{\varphi}_{c}\varepsilon_{,X}(t_{c})=-\varepsilon_{,\varphi}(t_{c}).\label{eq:eq at tcrit}\end{equation}
As follows from Eq.~ (\ref{eq:energy density on X}), $\varepsilon_{,X}(t_{c})=p_{,X}(t_{c})$
at the time $t_{c}$ when the system crosses $X_{c}=0$. Provided
$\varepsilon_{,\varphi}(t_{c})\neq0$ and $p_{,X}(t_{c})\neq0$, we
infer from Eq.~(\ref{eq:w two dot}) and (\ref{eq:eq at tcrit})
that the equation of state parameter $w(t)$ has either a minimum
or a maximum at the point $t_{c}$. Thus, the transition is impossible
in this case. 

If $\varepsilon_{,\varphi}(t_{c})=0$ and $\varepsilon_{,X}(t_{c})\neq0$,
then it follows from relation (\ref{eq:eq at tcrit}) that $\ddot{\varphi}(t_{c})=0$.
Therefore the considered solution $\varphi(t)$ for which $w\left[\varphi(t_{c})\dot{\varphi}(t_{c})\right]=-1$
is a fixed point solution $\varphi(t)\equiv\varphi_{c}\equiv const$
and the transition is impossible. Since $\varepsilon>0$, we see that
this fixed point is obviously the de Sitter solution. 

If not only $X_{c}=0$ but also $\varepsilon_{,X}(t_{c})\equiv p_{,X}(t_{c})=0$,
then $\varepsilon_{,\varphi}(t_{c})=0$ and it follows from the formula
(\ref{eq:energy density on X}) that $p_{,\varphi}(t_{c})=0$. Moreover,
the equation of motion (\ref{eq:Eq. of moution}) is not solved with
respect to the highest derivatives (namely, with respect to $\ddot{\varphi}$
) and therefore does not necessarily have a unique solution. It happens
because the point $(\varphi_{c},0)$ on the phase plot $(\varphi,\dot{\varphi})$
does not determine the $\ddot{\varphi}$ via the equation of motion
(\ref{eq:Eq. of moution}). It is clear that, in this case, the pointlike
(on the phase plot) solution $\varphi(t)\equiv\varphi_{c}\equiv const$
is a solution, but not necessary a unique one.

\begin{figure}
\begin{center}\includegraphics[%
  width=3.3in]{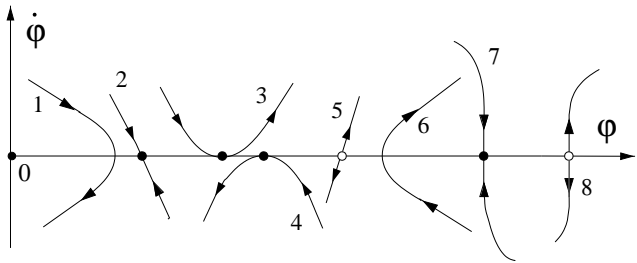}\end{center}

\caption{\label{cap:Phase-curves-in}Possible phase curves in the neighborhood
of the $\varphi-$axis. Only on the curves 1 and 6, the system crosses
the $\varphi$ axis. Curves 2, 3, 4 and 7 have an attractor as a shared
point with the $\varphi$ axis, whereas curves 5 and 8 have a repulsor.
These attractors and repulsors can be fixed-point solutions or singularities. }
\end{figure}

Below, we will give a more general consideration of the geometry of
phase curves in the neighborhood of the $\varphi$ axis. The phase
flows are directed from right to left for the lower part of the phase
plot $\dot{\varphi}<0$ and from left to right for the upper part
$\dot{\varphi}>0$, see Fig~\ref{cap:Phase-curves-in}.%
 Therefore the system can pass the $\varphi$ axis only if the point
of intersection is a turning point (curves 1 and 6 on Fig~\ref{cap:Phase-curves-in}).
Otherwise the crossing is a fixed point (or a singularity). If there
is a smooth phase curve on which the system pass through the $\varphi$
axis, then in a sufficiently small neighborhood of the turning point
we have $\left(\varphi-\varphi_{c}\right)\sim\dot{\varphi}^{2n}\sim X^{n}$
where $n\geq1$. Restricted to this curve, the function $Xp_{,X}(\varphi,X)$
depends only on $X$ and in the absence of a branching point the sign
of this function above and below the $\varphi$ axis is the same.
Then it follows from the formula (\ref{eq:W on fi dot}) that the
system cannot change the sign of $\left(w+1\right)$ while crossing
the $\varphi$ axis. 

If a smooth phase curve does not cross but touches the $\varphi$
axis at a point $\varphi_{c}$ (see Fig.~\ref{cap:Phase-curves-in},
trajectories 3 and 4), then the following asymptotic holds: $\dot{\varphi}\sim\left(\varphi-\varphi_{c}\right)^{2n}$,
where $n\geq1$. Let us find the time needed for the system to reach
the tangent point $(\varphi_{c},0)$ in this case. We have \begin{equation}
t\equiv\int_{\varphi_{in}}^{\varphi_{c}}\frac{d\varphi}{\dot{\varphi}(\varphi)}\sim\int_{\varphi_{in}}^{\varphi_{c}}\frac{d\varphi}{\left(\varphi-\varphi_{c}\right)^{2n}},\end{equation}
where $\varphi_{in}$ is a starting point on the phase curve. The
last integral is obviously divergent. Therefore the system cannot
reach the tangent point in a finite time 

Finally, we come to the conclusion that in the framework under consideration
it is impossible to build a model with the desirable transition through
the points $X_{c}=0$.

\subsection{\label{sub:Transition-at-C^2=3D0}Transition at points $\Psi_{c}$:
$p_{,X}(\Psi_{c})=0$, $\varepsilon_{,X}(\Psi_{c})\neq0$,~$X_{c}\neq0$ }

In the neighborhood of a point $\Psi_{c}$, at which the condition
$\varepsilon_{,X}(\Psi_{c})\neq0$ holds, one can find a function~$\dot{\varphi}_{c}(\varphi_{c})$:
$p_{,X}\left[\varphi_{c},\dot{\varphi}_{c}(\varphi_{c})\right]=0$.
This follows from the theorem about the implicit function. One would
anticipate that on the phase curves intersecting the curve $\dot{\varphi}_{c}(\varphi_{c})$
the state of the dark energy changes to the phantom one (or vise versa).
Let us express $p_{,X}$ from Eq.~(\ref{eq:W on fi dot}) and substitute
it into formula (\ref{eq:saund speed}) for the sound speed of perturbations:\begin{equation}
c_{s}^{2}=\frac{(w+1)\varepsilon}{2X\varepsilon_{,X}}.\label{eq:Csound  on W}\end{equation}
 For the stability with respect to the general metric and matter perturbations
the condition $c_{s}^{2}\geq0$ is necessary (see \cite{Perturbations}).
Indeed the increment of instability is inversely proportional to the
wavelength of the perturbations, and hence the background models for
that $c_{s}^{2}<0$ are violently unstable and do not have any physical
significance. Because of the continuity of $\varepsilon_{,X}$, there
exists a neighborhood of the point $\Psi_{c}$ where $\varepsilon_{,X}\neq0$.
Therefore, from the above expression for the sound speed (\ref{eq:Csound  on W})
it follows that if $\left(w+1\right)$ change a sign then $c_{s}^{2}$
should change a sign as well. If this is the case, then the trajectories,
realizing the transition, violate the stability condition $c_{s}^{2}\geq0$.
Therefore the model of the transition is not realistic.

\subsection{\label{sub:Transition-at-points}Transition at points $\Psi_{c}$:
$p_{,X}(\Psi_{c})=0$, $\varepsilon_{,X}(\Psi_{c})=0$,~$X_{c}\neq0$ }

As we have already mentioned at the beginning of this section, the
points $\Psi_{c}$ generally form the curves in the phase space $(\varphi,\dot{\varphi})$.
The subclass of the points $\Psi_{c}$, which we are going to consider
in this subsection, is generally a collection of the isolated points
given by the solutions of the system,\begin{eqnarray}
p_{,X}(\varphi,\dot{\varphi})=0, &  & \varepsilon_{,X}(\varphi,\dot{\varphi})=0.\end{eqnarray}
 Only for specific models, the solutions of this system are not isolated
points. An example when these solutions form a line is considered
in section \ref{Examples}. Usually the phase curves passing through
the isolated points build a set of the zero measure. Therefore it
is physically implausible to observe the processes realized on these
solutions. The only reason to study the behaviour of the system around
these points is their singular character. The point is that the equation
of motion (\ref{eq:Eq. of moution}) is not solved with respect to
the highest derivatives at this points. In such points there can be
more than one phase curve passing through each point. Moreover, the
set of solutions $\varphi(t)$, which pass through $\Psi_{c}$ with
different $\ddot{\varphi}$, could have a non-zero measure. On the
other hand, the equation of motion does not necessarily have a solution
$\varphi(t)$ such that $(\varphi(t_{c}),\dot{\varphi}(t_{c}))=\Psi_{c}$
at some moment of time $t_{c}$, or there exists the desirable solution
$\varphi(t)$ but it does not possess the second derivative with respect
to time at the point $\Psi_{c}$. Below we will analyze the behavior
of the phase curves in the neighborhoods of the points $\Psi_{c}$.

The equation of motion (\ref{eq:Eq. of moution}) can be rewritten
as a system of two differential equations of the first order:\begin{eqnarray}
 &  & \frac{d\dot{\varphi}}{dt}=-\dot{\varphi}\frac{p_{,X}}{\varepsilon_{,X}}\sqrt{3\varepsilon}-\frac{\varepsilon_{,\varphi}}{\varepsilon_{,X}},\label{eq:system of first order equation}\\
 &  & \frac{d\varphi}{dt}=\dot{\varphi}.\nonumber \end{eqnarray}
 The phase curves of this dynamical system are given by the following
differential equation:\begin{equation}
\frac{d\dot{\varphi}}{d\varphi}=-\frac{\dot{\varphi}p_{,X}\sqrt{3\varepsilon}+\varepsilon_{,\varphi}}{\dot{\varphi}\varepsilon_{,X}}.\label{eq:Trajectory}\end{equation}
This equation follows from the system (\ref{eq:system of first order equation})
and therefore all phase curves corresponding to the integral curves
of system (\ref{eq:system of first order equation}) are integral
curves of the differential equation (\ref{eq:Trajectory}). But the
reverse statement is false, so each integral curve of Eq.~(\ref{eq:Trajectory})
does not necessarily correspond to a solution $\varphi(t)$ of the
equation of motion (or of the system (\ref{eq:system of first order equation})).
In the neighborhoods of the points where $\varepsilon_{,X}\neq0$,
it is convenient to introduce a new auxiliary time variable $\tau$
defined by \begin{equation}
dt\equiv\varepsilon_{,X}d\tau.\label{eq:dt}\end{equation}
The system (\ref{eq:system of first order equation}) is equivalent
to the $\tau$ system: \begin{eqnarray}
 &  & \frac{d\dot{\varphi}}{d\tau}=-\dot{\varphi}p_{,X}\sqrt{3\varepsilon}-\varepsilon_{,\varphi},\label{eq:tau-system}\\
 &  & \frac{d\varphi}{d\tau}=\dot{\varphi}\varepsilon_{,X}.\nonumber \end{eqnarray}
The auxiliary time variable $\tau$ change the direction if $\varepsilon_{,X}$
change the sign. Note that the system (\ref{eq:tau-system}) always
possesses the same phase curves as the equation of motion (\ref{eq:system of first order equation}). 

\begin{figure}
\begin{center}\includegraphics[%
  width=3in]{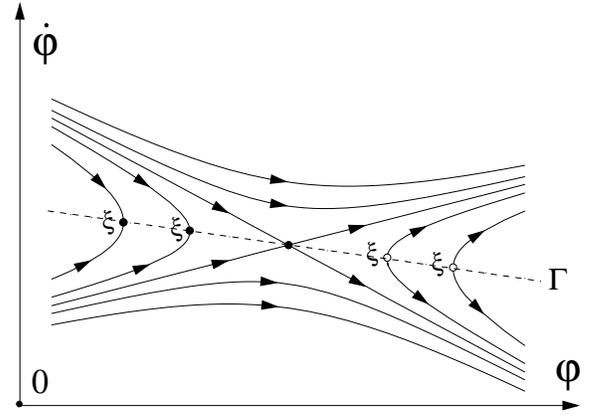}\end{center}

\caption{\label{Saddle}Phase curves in the neighborhood of the singular point
$\Psi_{c}^{+}$ are plotted for the case of the real $\lambda$. At
the points $\xi$, the solutions $\varphi(t)$ do not exist. These
points together with $\Psi_{c}^{+}$ form the curve $\Gamma$ on which
$\varepsilon_{,X}(\Gamma)=0$.}
\end{figure}

\begin{figure}
\begin{center}\includegraphics[%
  width=3in]{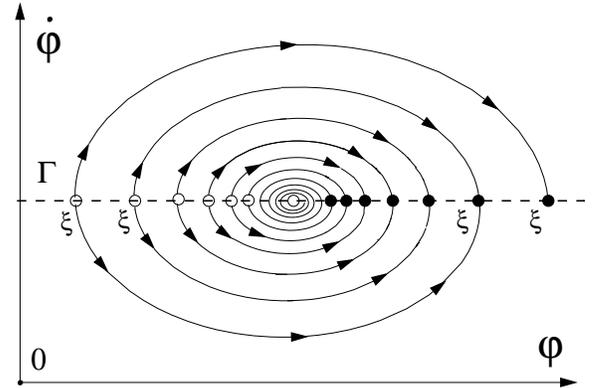}\end{center}

\caption{\label{cap:Focus} Phase curves in the neighborhood of the singular
point $\Psi_{c}^{+}$ are plotted for the case of the pure imaginary
$\lambda$. Here we assume that the singular point is a focus.}
\end{figure}

In the case under consideration we have $\dot{\varphi}_{c}\neq0$
and from the formula\begin{equation}
\frac{d\dot{\varphi}}{d\varphi}=\frac{\ddot{\varphi}}{\dot{\varphi}}\end{equation}
we infer that $d\dot{\varphi}/d\varphi(t_{c})$ should be finite,
if $\ddot{\varphi}(t_{c})$ is finite. As one can see from the equation
determining the phase curves (\ref{eq:Trajectory}), in order to obtain
a finite $d\dot{\varphi}/d\varphi(t_{c})$ it is necessary that at
least $\varepsilon_{,\varphi}(\Psi_{c})=0$. In the case if $\varepsilon_{,\varphi}(\Psi_{c})\neq0$,
the solution $\varphi(t)$ does not possess the second $t$ derivative
at the point $t_{c}$. Usually this can be seen as unphysical situation.
But nevertheless this does not necessarily lead to the unphysical
incontinuity in the observed quantities $\varepsilon,$ $p$, $H$,
and $\varphi$, $\dot{\varphi}$. One may probably face problems with
the stability of such solutions, but let us first of all investigate
the behavior of the phase curves in the case $\varepsilon_{,\varphi}(\Psi_{c})\neq0$.
From Eq.~(\ref{eq:Trajectory}), we obtain $d\varphi/d\dot{\varphi}=0$
at $\Psi_{c}$. Further we can parameterize the phase curve as $\varphi=\varphi(\dot{\varphi})$
and bring the equation for phase curves (\ref{eq:Trajectory}) to
the form \begin{equation}
\frac{d\varphi}{d\dot{\varphi}}=\varepsilon_{,X}(\varphi,\dot{\varphi})F(\varphi,\dot{\varphi}),\end{equation}
where we denote\begin{equation}
F(\varphi,\dot{\varphi})\equiv-\frac{\dot{\varphi}}{\dot{\varphi}p_{,X}\sqrt{3\varepsilon}+\varepsilon_{,\varphi}}.\end{equation}
 If, as we have assumed, $\varepsilon_{,\varphi}(\Psi_{c})\neq0$
and $\varepsilon(\Psi_{c})\neq0$, then $F(\varphi,\dot{\varphi})$
is differentiable in the neighborhood of the point $\Psi_{c}$ and
$F(\Psi_{c})=-\dot{\varphi}/\varepsilon_{,\varphi}$. For the second
$\dot{\varphi}$ derivative at the point $\Psi_{c}$, one obtains\begin{equation}
\frac{d^{2}\varphi}{d^{2}\dot{\varphi}}=-\frac{\dot{\varphi}^{2}}{\varepsilon_{,\varphi}}\varepsilon_{,XX}.\end{equation}
 That is why the point $\Psi_{c}$ is a minimum or a maximum for the
function $\varphi(\dot{\varphi})$. In this case $\Psi_{c}$ is such
an exceptional point on the phase plot, where the solution $\varphi(t)$
cannot have continuous $\dot{\varphi}(t)$ and the phase curve terminates
(see points $\xi$ in Figs.~\ref{Saddle} and \ref{cap:Focus}).
This happens because the direction of the phase flow is preserved
in the neighborhood of $\Psi_{c}$. If $\varepsilon_{,XX}(\Psi_{c})=0$,
then one can find the third $\dot{\varphi}$ derivative of $\varphi(\dot{\varphi})$
at the point $\Psi_{c}$:\begin{equation}
\frac{d^{3}\varphi}{d^{3}\dot{\varphi}}=-\frac{\dot{\varphi}^{3}}{\varepsilon_{,\varphi}}\varepsilon_{,XXX}.\end{equation}
 In this case there can exist a continuous solution $\varphi(t)$
such that $(\varphi(t_{c}),\dot{\varphi}(t_{c}))=\Psi_{c}$ at some
moment of time $t_{c}$ and the only bad thing happening in this point
is that $\ddot{\varphi}(t_{c})$ does not exist. Let us now investigate
what happens with the equation of state at this point of time. Differentiating
both sides of the definition (\ref{eq:W for P}) of $w$ yields at
$t_{c}$\begin{equation}
\dot{w}_{c}=\left[\frac{\dot{\varphi}}{\varepsilon}\left(p_{,\varphi}-c_{s}^{2}\varepsilon_{,\varphi}\right)\right]_{c},\end{equation}
 where we have used the equation of motion (\ref{eq:Eq. of moution})
at the point $t_{c}$ and the definition (\ref{eq:saund speed}) of
$c_{s}^{2}$. Applying the l'H\^opital rule for the $c_{s}^{2}(t_{c})=\lim\limits _{t\rightarrow t_{c}}p_{,X}/\varepsilon_{,X}$,
we find that $c_{s}^{2}(t_{c})=0$. Moreover, using the l'H\^opital
rule for the derivative of $c_{s}^{2}$ at the point $\Psi_{c}$,
one can find that $dc_{s}^{2}/d\dot{\varphi}\sim p_{,\varphi}/\varepsilon_{,\varphi}$.
Thus, if $p_{,\varphi}\neq0$ the transition could occur but it changes
a sign of the sound speed $c_{s}^{2}$. Therefore, if the stability
criteria are applicable to this case, then the transition leads to
instability. \\

\begin{figure}
\begin{center}\includegraphics[%
  width=3in]{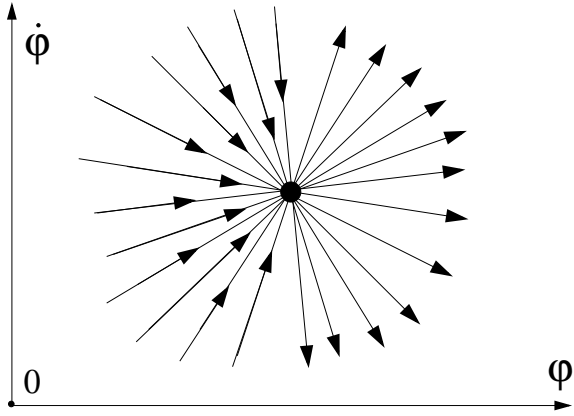}\end{center}

\caption{\label{cap:Node}If $\mathbf{A}$ had the eigenvalues $\lambda_{1}=\lambda_{2}$,
then the singular point $\Psi_{c}^{+}$ would be a nodal point and
there would be a continuous set of trajectories passing through it.
To illustrate this we plot here the phase curves in the particular
case of a degenerate nodal point. The form of the equation of motion
(\ref{eq:Eq. of moution}) excludes such types of singular points
and therefore prevents the possibility of such transitions.}
\end{figure}

The necessary condition for the existence of $\ddot{\varphi}$ during
the transition is\begin{equation}
\varepsilon_{,\varphi}(\Psi_{c})=2X_{c}p_{,X\varphi}(\Psi_{c})-p_{,\varphi}(\Psi_{c})=0.\label{eq:X-Condition}\end{equation}
This condition drastically reduces the set of the points $\Psi_{c}$,
where the transition is possible. Namely, they are the critical points
of the function $\varepsilon(\varphi,X)$ and, on the other hand,
they are the fixed points of the auxiliary $\tau$ system (\ref{eq:tau-system}).
These fixed points are additional to the fixed points of the system
(\ref{eq:system of first order equation}) defined by $\dot{\varphi}=0$
, $\varepsilon_{,\varphi}(\varphi,0)=0$, and $\varepsilon_{,X}(\varphi,0)\neq0$.
From now on, we will consider only those points $\Psi_{c}^{+}$ where
the condition (\ref{eq:X-Condition}) holds. From the relation (\ref{eq:X-Condition}),
it follows that if $p_{,\varphi}(\varphi_{c},X_{c})=0$ then $p_{,X\varphi}(\varphi_{c},X_{c})=0$.
Otherwise $X_{c}=0$, and as we have already seen the transition cannot
happen via the points $X_{c}=0$. Note that if $p_{,\varphi}(\Psi_{c}^{+})=0$,
then the points $\Psi_{c}^{+}$ are common critical points of the
pressure $p(\varphi,\dot{\varphi})$, energy density $\varepsilon(\varphi,\dot{\varphi})$,
and $p_{,X}(\varphi,\dot{\varphi})$. From condition (\ref{eq:X-Condition})
follows that points $\Psi_{c}^{+}$ are singular points of Eq.~(\ref{eq:Trajectory}).
In such points there can be more than one phase curve passing through
this point. Moreover, as we have already mentioned, the set of solutions
$\varphi(t)$, which pass through $\Psi_{c}^{+}$ with different $\ddot{\varphi}$,
could have a nonzero measure. For example, if $\Psi_{c}^{+}$ were
a nodal point (see Fig.~\ref{cap:Node}), there would be a continuous
amount of the solutions passing through this point and therefore there
would be a continuous amount of solutions on which the transition
could occur. \\

Let us investigate the type of the singular points $\Psi_{c}^{+}$.
This will tell us about the amount of the solutions $\varphi(t)$
on which the transition is possible and their stability. For this
analysis, one can use the technique described, for example, in \cite{Stepanov},
and consider the integral curves of the equation (\ref{eq:Trajectory}).
Here we proceed with this analysis in a more convenient way, namely,
using the auxiliary $\tau$ system (\ref{eq:tau-system}). It is convenient
because for this system the singular points $\Psi_{c}^{+}$ are usual
fixed points. As we have already mentioned, both systems have the
same phase curves and therefore the analysis to perform is also applicable
to the phase curves of the system (\ref{eq:system of first order equation}).
The only thing we should not forget is the difference in the directions
of the phase flows of these systems. If $\varepsilon(\Psi_{c}^{+})\neq0$,
then one can linearize the right-hand side of the $\tau$ system in
the neighborhood of a point $\Psi_{c}^{+}$: $(\varphi_{c}^{+}+\delta\varphi,\dot{\varphi}_{c}^{+}+\delta\dot{\varphi})$.
The linearized $\tau$ system (\ref{eq:tau-system}) is \begin{equation}
\frac{d}{d\tau}\mathbf{V}=\mathbf{AV},\label{eq:linarized trajectory genral}\end{equation}
where we denote \begin{eqnarray}
\mathbf{V=\left(\mathbf{\begin{array}{c}
\delta\varphi\\
\delta\dot{\varphi}\end{array}}\right)}, &  & \mathbf{A=\left(\begin{array}{cc}
a & b\\
c & g\end{array}\right)},\label{eq:matrix A}\end{eqnarray}
and elements of the matrix $\mathbf{A}$ are given by the formulas\begin{align}
 & a=\varepsilon_{,X\varphi}\dot{\varphi},\label{eq:coeficients}\\
 & b=2X\varepsilon_{,XX},\nonumber \\
 & c=-\left(3H\dot{\varphi}p_{,X\varphi}+\varepsilon_{,\varphi\varphi}\right),\nonumber \\
 & g=-\varepsilon_{,X\varphi}\dot{\varphi},\nonumber \end{align}
where all quantities are calculated at $\Psi_{c}^{+}$. Here we have
used the Friedmann equation (\ref{Einstein's equation 2}). If $\Psi_{c}^{+}$
is an isolated fixed point of the $\tau$ system (\ref{eq:tau-system})
(or equivalently the singular point of system (\ref{eq:system of first order equation})),
then the following condition holds\begin{equation}
\det\mathbf{A}=ag-bc\neq0.\end{equation}
 The type of the fixed point depends on the eigenvalues $\lambda$
of the matrix $\mathbf{A}$ (for details see, for example, \cite{Pontryagin}).
In the case under consideration $a=-g$ and therefore we have \begin{equation}
\lambda^{2}=bc+a^{2}=-\det\mathbf{A}.\label{eq:root}\end{equation}
If $bc+a^{2}>0$, then eigenvalues $\lambda$ are real and of the
opposite signs. In accordance with the classification of the singular
points, $\Psi_{c}^{+}$ is a saddle point (see Fig.~\ref{Saddle}).
Therefore the transition is absolutely unstable; there are only two
solutions $\varphi(t)$ on which the transition is allowed to occur.\\

If $bc+a^{2}<0$ then $\lambda$ are pure imaginary. Here the situation
is a little bit more complicated: In accordance with \cite{Stepanov},
this fixed point of a nonlinear system can be either a focus or a
center. In these cases, as one can see from Fig.~\ref{cap:Focus},
there no solutions $\varphi(t)$ passing through the point $\Psi_{c}^{+}$.
Therefore, from now on we will consider only the first case - real
$\lambda$ of the opposite signs. 

It is convenient to rewrite the expression for $\lambda$ into a simpler
form. Differentiating the continuity equation (\ref{eq:continuity})
yields\begin{equation}
\ddot{\varepsilon}_{c}=-3H_{c}\dot{p}_{c}.\end{equation}
Remember that the index $c$ denotes quantities taken at $t_{c}$
or in this subsection at $\Psi_{c}^{+}$. Differentiating the pressure
$p$ as a composite function, we have $\dot{p}=p_{,\varphi}\dot{\varphi}+p_{,X}\dot{X}$.
Assuming that $\ddot{\varphi}_{c}$ is finite, we obtain that $\dot{p}_{c}=p_{,\varphi}^{c}\dot{\varphi}_{c}$.
Thus the formula for $\ddot{\varepsilon}_{c}$ is\begin{equation}
\ddot{\varepsilon}_{c}=-3H_{c}p_{,\varphi}^{c}\dot{\varphi}_{c}.\label{eq:e second derivative}\end{equation}
Using the condition (\ref{eq:X-Condition}) and the last equation,
we bring the element $c$ of the matrix $\mathbf{A}$ to the following
form:\begin{equation}
c=-\varepsilon_{,\varphi\varphi}^{c}+\frac{\ddot{\varepsilon}_{c}}{2X_{c}}.\end{equation}
 This relation allows one to rewrite the formula (\ref{eq:root})
as follows:\begin{equation}
\lambda^{2}=2X_{c}\left(\left(\varepsilon_{,X\varphi}^{c}\right)^{2}-\varepsilon_{,\varphi\varphi}^{c}\varepsilon_{,XX}^{c}\right)+\ddot{\varepsilon}_{c}\varepsilon_{,XX}^{c}.\label{eq:lamda}\end{equation}
 Here it is interesting to note that the expression $\varepsilon_{,\varphi\varphi}^{c}\varepsilon_{,XX}^{c}-\left(\varepsilon_{,X\varphi}^{c}\right)^{2}$
from the previous formula is the determinant of the quadratic form
arising in the Taylor set of $\varepsilon$ in the neighborhood of
the critical point $\Psi_{c}^{+}$. If this determinant is positive,
then the function $\varepsilon(\varphi,X)$ has either a minimum or
a maximum at point $\Psi_{c}^{+}$. Otherwise there is either one
curve of constant energy density $\varepsilon(\textrm{curve})=\varepsilon(\Psi_{c}^{+})$
with a singular turning point at $\Psi_{c}^{+}$ in the neighborhood
of $\Psi_{c}^{+}$ ore two intersecting at $\Psi_{c}^{+}$ curves
of constant $\varepsilon$ (see \cite{Nikolsky}) . On the other hand,
differentiating $\varepsilon$ as a composite function we find\begin{equation}
\ddot{\varepsilon}_{c}=2X_{c}\left(\varepsilon_{,\varphi\varphi}^{c}+2\varepsilon_{,X\varphi}^{c}\ddot{\varphi}_{c}+\varepsilon_{,XX}^{c}\ddot{\varphi}_{c}^{2}\right).\label{eq:e second derivative two}\end{equation}
Substituting this relation into the previous formula (\ref{eq:lamda})
for $\lambda$ yields \begin{equation}
\lambda^{2}=2X_{c}\left(\varepsilon_{,XX}^{c}\ddot{\varphi}_{c}+\varepsilon_{,X\varphi}^{c}\right)^{2}.\label{eq:lamda final}\end{equation}
 This formula provides the relation between $\ddot{\varphi}_{c}$
at the moment of transition and $\lambda$. Note that $\lambda$ depends
on $\ddot{\varphi}_{c}$ only in the case when $\varepsilon_{,XX}(t_{c})\equiv2X_{c}p_{,XXX}(t_{c})\neq0$.
Moreover, comparing formulas (\ref{eq:e second derivative}) and (\ref{eq:e second derivative two})
for $\ddot{\varepsilon}_{c}$, one can obtain the equation on $\ddot{\varphi}_{c}$:\begin{equation}
\varepsilon_{,XX}^{c}\dot{\varphi}_{c}\ddot{\varphi}_{c}^{2}+2\varepsilon_{,X\varphi}^{c}\dot{\varphi}_{c}\ddot{\varphi}_{c}+\dot{\varphi}_{c}\varepsilon_{,\varphi\varphi}^{c}+3H_{c}p_{,\varphi}^{c}=0.\label{eq:quadrat eq on fi two dot}\end{equation}
This equation is solvable in real numbers if the discriminant is positive.
As one can prove, the discriminant is exactly the $4\lambda^{2}$
and therefore positive if we consider the saddle point. The same can
be seen from relation (\ref{eq:lamda final}) as well. 

Let us denote the positive and negative eigenvalues and the corresponding
eigenvectors of $\mathbf{A}$ as $\lambda_{+}$, $\lambda_{-}=-\lambda_{+}$
and $\mathbf{a}_{+}$, $\mathbf{a}_{-}$, respectively. 

If $b\neq0$ ( or equivalently $\varepsilon_{,XX}^{c}\neq0$), then
the eigenvectors can be chosen as $\mathbf{a}_{+}=(1,(\lambda_{+}-a)/b)$
and $\mathbf{a}_{-}=(1,-(\lambda_{+}+a)/b)$. Therefore the separatrices
forming the saddle are \begin{eqnarray*}
\delta\dot{\varphi}_{+}=\frac{\lambda_{+}-a}{b}\delta\varphi & \textrm{and} & \delta\dot{\varphi}_{-}=-\frac{\lambda_{+}+a}{b}\delta\varphi.\end{eqnarray*}
The general solution for the phase curves in the neighborhood of $\Psi_{c}^{+}$
is\begin{equation}
\left(\delta\dot{\varphi}-\frac{\lambda_{+}-a}{b}\delta\varphi\right)\left(\delta\dot{\varphi}+\frac{\lambda_{+}+a}{b}\delta\varphi\right)=const.\label{eq:trajectory exxnonzero}\end{equation}
If $b=0$ and additionally $a>0$, then we have $\lambda_{+}=a=\varepsilon_{,X\varphi}\dot{\varphi}$,
and one can choose the eigenvectors as $\mathbf{a}_{+}=(1,c/2a)$
and $\mathbf{a}_{-}=(0,1)$ . For a negative $a$, one can obtain
the eigenvectors and eigenvalues by changing $\lambda_{+}\leftrightarrow\lambda_{-}$
and $\mathbf{a}_{+}\leftrightarrow\mathbf{a}_{-}$. The separatrices
then read \begin{eqnarray*}
\delta\varphi=0 & \textrm{and} & \delta\dot{\varphi}=\frac{c}{2a}\delta\varphi.\end{eqnarray*}
Thus, similarly to the previous case, the phase curves are given by\begin{equation}
\delta\varphi\left(\delta\dot{\varphi}-\frac{c}{2a}\delta\varphi\right)=const.\label{eq:trajectories for zero exx}\end{equation}
As we have already mentioned from the formula $d\dot{\varphi}/d\varphi$$=\ddot{\varphi}/\dot{\varphi}$,
follows that, at the points where the phase curves are parallel to
the $\dot{\varphi}$ axis and where $\dot{\varphi}\neq0$, the second
$t$ derivative of the field does not exist. If we look at the equations
(\ref{eq:trajectory exxnonzero}) and (\ref{eq:trajectories for zero exx})
providing the phase curves in the neighborhood of $\Psi_{c}^{+}$,
then we find that in the first case (when $\varepsilon_{,XX}^{c}\neq0$)
the phase curves lying on the right- and left- hand sides of both
separatrices should have a point $\xi$ where they are parallel to
the $\dot{\varphi}$ axis (see Fig.~\ref{Saddle}). Therefore each
of these phase curves consists of two solutions of the equation of
motion (\ref{eq:Eq. of moution}) and the exceptional point $\xi$
where the solution $\varphi(t)$ does not exist. The same statement
holds in the case of the pure imaginary $\lambda$ (see Fig.~\ref{cap:Focus}).
This behavior is not forbidden because, as one can easily prove, the
exceptional points $\xi$ lie exactly on the curve $\Gamma$ on which
$\varepsilon_{,X}(\Gamma)=0$ and the equation of motion is not solved
with respect to the highest derivatives. Note that we have already
assumed $\lambda\neq0$, and from this condition it follows that $\varepsilon_{,XX}^{c}$
and $\varepsilon_{,X\varphi}^{c}$ cannot vanish simultaneously. Therefore
in the neighborhood of the point $\Psi_{c}^{+}$ there exists an implicit
function $\dot{\varphi}(\varphi)$ (or $\varphi(\dot{\varphi})$)
and its plot gives the above-mentioned curve $\Gamma$ on which $\varepsilon_{,X}(\Gamma)=0$.
In the case $\varepsilon_{,XX}^{c}=0$, the separatrix $\delta\varphi=0$
locally coincides with $\Gamma$, and therefore, this integral curve
of Eq.~(\ref{eq:Trajectory}) does not correspond to any solution
$\varphi(t)$ of the equation of motion. Nevertheless, in virtue of
the existence theorem, all phase curves obtained in the neighborhood
of the separatrix $\delta\varphi=0$ correspond to the solutions of
the equation of motion. Moreover, if $\varepsilon_{,XX}^{c}=0$ and
$p_{,X\varphi}^{c}\neq0$ then the curves on which $p_{,X}=0$ and
$\varepsilon_{,X}=0$ locally coincide with each other and with the
curve $\delta\varphi=0$. The only phase curve intersecting the curve
$\delta\varphi=0$ at $\Psi_{c}^{+}$ is the second separatrix $\delta\dot{\varphi}=c\delta\varphi/2a$.
Thus the only solution $\varphi(t)$ on which the transition happens
in the neighborhood of $\Psi_{c}^{+}$ corresponds to the separatrix
$\delta\dot{\varphi}=c\delta\varphi/2a$. This can also be seen from
the equation (\ref{eq:quadrat eq on fi two dot}) which has only one
root $\ddot{\varphi}_{c}$ in this case. In the Sec.~(\ref{Examples})
we will illustrate this with a numerical example (see Fig.~\ref{cap:uplus}).\\

It is worthwhile to discuss cases which fall out from the consideration
made above. We have assumed that $\lambda\neq0$ and therefore $\Psi_{c}^{+}$
is an isolated singular point of Eq.~(\ref{eq:Trajectory}). The
most natural possibilities to drop out this condition are $\varepsilon_{,X\varphi}^{c}=0$
and either $\varepsilon_{,XX}^{c}=0$ or $\dot{\varphi}_{c}3H_{c}p_{,X\varphi}^{c}+\varepsilon_{,\varphi\varphi}^{c}=0$.
In the first case $\Psi_{c}^{+}$ is a critical point not only of
the function $\varepsilon$ but of the function $\varepsilon_{,X}$
as well. This can be obtained either for a very special kind of function
$p$ namely, such that $p_{,X}=0$, $p_{,XX}=0$, $p_{,XXX}=0$, and
$\left[p-4X^{2}p_{,XX}\right]_{,\varphi}=0$ at $\Psi_{c}^{+}$ or
imposing the condition that the point $\Psi_{c}^{+}$ is a critical
point not only of $p$ but also of the functions $p_{,X}$ and $p_{,XX}$:
$p_{,X}=0$, $p_{,\varphi}=0$, $p_{,XX}=0,$ $p_{,X\varphi}=0$,
$p_{,XX\phi}=0$, and finally $p_{,XXX}=0$ at $\Psi_{c}^{+}$. In
the second case $\Psi_{c}^{+}$ is a common critical point for the
functions $\varepsilon$ and $\varepsilon_{,\varphi}$. In terms of
$p$, this condition is as follows: $p_{,X}=0$, $p_{,\varphi}=0$,
$p_{,X\varphi}=0$, $p_{,XX}=0,$ $p_{,\varphi\varphi}=0$, $p_{,XX\varphi}=0$,
and finally $p_{,X\varphi\varphi}=0$ at $\Psi_{c}^{+}$ . Thus, the
point $\Psi_{c}^{+}$ is a common critical point of $p$, $p_{,X}$
and $p_{,X\varphi}$. Of course, the analysis performed above does
not work in the case if the function $p(\varphi,X)$ does not have
a sufficient amount of derivatives. It is clear that all these cases
are not general.\\

Let us sum up the results obtained in this section. In the general
case of linearizable functions $\varepsilon_{,X}$, $\varepsilon_{,\varphi},$
and $p_{,X}$, the considered transitions either occur through the
points $\Psi_{c}^{+}$, where $p_{,X}=0$, $\varepsilon_{,X}=0$,
$\varepsilon_{,\varphi}=0$, and \[
\dot{\varphi}\left[\left(\varepsilon_{,X\varphi}^{2}-\varepsilon_{,\varphi\varphi}\varepsilon_{,XX}\right)\dot{\varphi}-3H\varepsilon_{,XX}p_{,\varphi}\right]>0\]
or lead to an unacceptable instability with respect to the cosmological
perturbations of the background. The points $\Psi_{c}^{+}$ are critical
points of the energy density and are the singular points of the equation
of motion of the field $\varphi$ as well. These singular points are
saddle points and the transition is realized by the repulsive separatrix
solutions, which form the saddle. Therefore the measure of these solutions
is zero in the set of trajectories and the dynamical transitions from
the states where $w>-1$ to $w<-1$ or vise versa are physically implausible.

\section{\label{Examples}Lagrangians linear in $X$ }

The simplest class of models, for that one could anticipate the existence
of dynamical transitions, is the dark energy described by Lagrangians
$p(\varphi,X)$ linear in $X$: \begin{equation}
p(\varphi,X)=KX-\frac{1}{2}V\equiv\frac{1}{2}\left(K(\varphi)\nabla_{\mu}\varphi\nabla^{\mu}\varphi-V(\varphi)\right).\end{equation}
In the isotropic and homogeneous Friedmann universe, the Lagrangian
is then\begin{equation}
p(\varphi,\dot{\varphi})=\frac{1}{2}\left(K(\varphi)\dot{\varphi}^{2}-V(\varphi)\right).\label{eq:Lagrange}\end{equation}
For these models, we always have $c_{s}^{2}=1$ and therefore, as
follows from our analysis, the transitions could occur only through
the points where $\varepsilon_{,X}=0$. The energy density for this
model is \begin{equation}
\varepsilon(\varphi,\dot{\varphi})=\frac{1}{2}\left(K(\varphi)\dot{\varphi}^{2}+V(\varphi)\right).\end{equation}
If one takes $K(\varphi)\equiv1$, then the Lagrangian (\ref{eq:Lagrange})
is the usual Lagrangian density for scalar field with a self-interaction.
If we take $K(\varphi)\equiv-1$, then we obtain the so-called {}``Phantom
field'' from \cite{Caldwell Manace} and \cite{Hoffman}. The case
$K(\varphi)>0$ corresponds to $w\geq-1$, whereas $K(\varphi)<0$
corresponds to $w\leq-1$. The equation of motion (\ref{eq:Eq. of moution})
takes in our case the following form:\begin{equation}
\ddot{\varphi}K+\dot{\varphi}K\sqrt{3\varepsilon}+\varepsilon_{,\varphi}=0.\label{eq:fi brifly evolution}\end{equation}
 While the equation determining the phase curves (\ref{eq:Trajectory})
reads in this particular case\begin{equation}
\frac{d\dot{\varphi}}{d\varphi}+\sqrt{3\varepsilon}+\frac{1}{\dot{\varphi}K}\varepsilon_{,\varphi}=0.\label{eq:Phase curve eq.}\end{equation}
If $K(\varphi)$ is a sign-preserving function, one can redefine field
$\varphi$: $\sqrt{\left|K(\varphi)\right|}d\varphi=d\phi$ (see also
Ref.~\cite{Liddle}). The equation of motion for the new field $\phi$
can be obtained from Eq.~(\ref{eq:fi brifly evolution}), through
the formal substitutions $\varphi\rightarrow\phi$, $V(\varphi)\rightarrow\tilde{V}(\phi)\equiv V(\varphi(\phi))$,
and $K(\varphi)\rightarrow\pm1$, where the upper sign corresponds
to a positive $K(\varphi)$ and the lower one to a negative $K(\varphi)$.
After these substitutions, the equation of motion (\ref{eq:fi brifly evolution})
looks more conventionally \begin{equation}
\ddot{\phi}+\dot{\phi}\sqrt{\frac{3}{2}\left(\pm\dot{\phi}^{2}+\tilde{V}(\phi)\right)}\pm\frac{1}{2}\left(\frac{\partial\tilde{V}(\phi)}{\partial\phi}\right)=0.\label{eq:red. Motion}\end{equation}
Moreover, this equations is easier to dial with, because one can visualize
the dynamic determined by it, as 1D classical mechanics of a point
particle in a potential $\pm\tilde{V}(\phi)/2$ with a little bit
unusual friction force. If we were able to solve the equation of motion
(\ref{eq:red. Motion}) for all possible $\tilde{V}(\phi)$ and initial
data, we could solve the problem of cosmological evolution for all
linear in $X$ Lagrangians with sign-preserving $K(\varphi)$.

If the function $K(\varphi)$ is not sign-preserving, then at first
sight it seems that the dark energy, described by such a Lagrangian,
can realize the desirable transition. The function $K(\varphi)$ generally
can change the sign in two ways: In the continuous one, then the function
$K(\varphi)$ takes the value zero for some values of field $\varphi$
or in a discontinuous polelike way.

\subsection{\label{sub:Differentiable-function-}Linearizable $K(\varphi)$}

Without loss of generality, one can assume that $K(0)\equiv K_{c}=0$,
$K(\varphi)<0$ for the negative values of $\varphi$ and $K(\varphi)>0$
for $\varphi>0$. The line $\varphi=0$ on the phase plot $(\varphi,\dot{\varphi})$
we will call the {}``critical'' line for the given class of Lagrangians.
The phantom states $(\varphi,\dot{\varphi})$ of the scalar field
lie on the left-hand side, while the usual states with $w\geq-1$
are on the right-hand side of the critical line. If there exists a
solution $\varphi(t)$ whose phase curve passes through the {}``critical''
line, then the dark energy can change the sign of $\left(w+1\right)$
during the cosmological evolution. From now on, we will investigate
the behavior of the phase curves of the system in the neighborhood
of the critical line. \\

First of all, it is worth considering the functions $K(\varphi)$
such that $K_{c}^{\prime}>0$ (here we have denoted $K{}_{,\varphi}(0)\equiv K{}_{c}^{\prime}$),
because in this case we can directly apply the outcome of our previous
analysis made in the Sec.~\ref{sub:Tansition-at-X=3D0}. Condition
(\ref{eq:X-Condition}) is for the linear in $X$ Lagrangians as follows:
\begin{equation}
\dot{\varphi}_{c}^{2}K_{c}^{\prime}+V_{c}^{\prime}=0.\label{eq:equation on phi dot}\end{equation}
 As we have already assumed $K_{c}^{\prime}>0$, therefore, if $V_{c}^{\prime}>0$,
then, as follows from condition (\ref{eq:equation on phi dot}), there
are no twice differentiable solutions $\varphi(t)$ whose phase curves
would intersect or touch the critical line. Further (see formula (\ref{eq:integral})
and below) we will show that, for the linear in $X$ Lagrangians,
condition (\ref{eq:X-Condition}) (or in our case condition (\ref{eq:equation on phi dot})
is necessary not only for the existence of the second $t$ derivative
$\ddot{\varphi}$ at the point of intersection with the critical line
but for the existence of a solution $\varphi(t)$ at this point as
well. Thus, we come to the conclusion that, if $V_{c}^{\prime}>0$,
then two regions $\varphi<0$, and $\varphi>0$ on the phase plot
are not connected by any phase curves and accordingly the dark energy
does not change the sign of $\left(w+1\right)$ during the cosmological
evolution.

In the case $V_{c}^{\prime}<0$, we can solve Eq.~(\ref{eq:equation on phi dot})
with respect to $\dot{\varphi}_{c}$ : \begin{equation}
\dot{\varphi}_{c}=u_{\pm}\equiv\pm\sqrt{-\frac{V_{c}^{\prime}}{K_{c}^{\prime}}}.\label{eq:uplusminus}\end{equation}
The phase curves, lying in the neighborhoods of the singular points
$\Psi_{c}^{+}=(0,u_{\pm})$, are to obtain from the relation (\ref{eq:trajectories for zero exx}),
which gives:\begin{equation}
\varphi\left(\dot{\varphi}-u_{\pm}-\frac{A_{\pm}}{2}\varphi\right)=const,\label{eq:singular phase curves}\end{equation}
where \begin{equation}
A_{\pm}=-3H_{c}+\frac{V_{c}^{\prime}K_{c}^{\prime\prime}-V_{c}^{\prime\prime}K_{c}^{\prime}}{2u_{\pm}\left(K_{c}^{\prime}\right)^{2}}.\end{equation}
For each singular point $(0,u_{\pm})$, there is a corresponding solution
$\varphi_{\pm}(t)$ whose phase curve is the separatrix \begin{equation}
\dot{\varphi}_{\pm}=u_{\pm}+\frac{A_{\pm}}{2}\varphi,\end{equation}
which intersects the critical line. These phase curves correspond
to the $const=0$ in the right-hand side of Eq.~(\ref{eq:singular phase curves}).
Another curve, which corresponds to $const=0$ is $\varphi=0$. As
we have already mentioned at the end of the previous subsection, this
curve does not correspond to any solutions $\varphi(t)$ of the equation
of motion (\ref{eq:fi brifly evolution}).

\begin{figure}
\begin{center}\includegraphics[%
  width=3.1in]{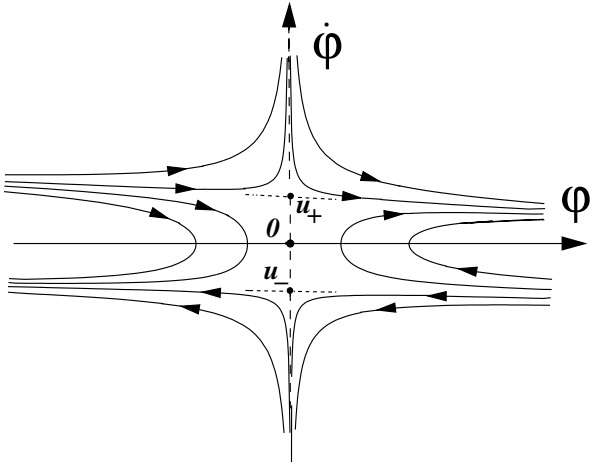}\end{center}

\caption{\label{cap:uplus}The typical behavior of the phase curves in the
neighborhood of the critical line where $K(\varphi)=0$ (here $\dot{\varphi}$
axis ) is plotted for the case when $K_{c}^{\prime}>0$ and $V_{c}^{\prime}<0$.
Horizontal dashed lines are the analytically obtained separatrices
$\dot{\varphi}_{\pm}$ and $(0,u_{\pm})$ are the points of transition.}
\end{figure}

Considering the phase flow in the neighborhoods of $\Psi_{c}^{+}=(0,u_{\pm})$
(see Fig.~\ref{cap:uplus}), we infer that the separatrices $\dot{\varphi}_{\pm}$
are repulsors immediately before they intersect the critical line
and attractors after the crossing. Hence, the measure of the initial
conditions $(\varphi,\dot{\varphi})$ leading to the transition to
phantom field (or vice versa) is zero. In this sense the dark energy
cannot change the sign of $K(\varphi)$ (or equivalently the sign
of $\left(w+1\right)$) during the cosmological evolution. 

The typical behavior of the phase curves in the neighborhood of the
singular points $(0,\, u_{\pm})$, for the models under consideration
($K_{c}^{\prime}>0$, $V_{c}^{\prime}<0$), is shown in Fig.~\ref{cap:uplus}.
Here, as an example, we have plotted the phase curves obtained numerically
for a toy model with the Lagrangian density $p=\frac{1}{2}\varphi\dot{\varphi}^{2}-\frac{1}{2}\left(\left(\varphi-1\right)^{2}+\frac{1}{3}\right)$.
For this model we have $u_{\pm}=\pm1$, $A_{+}=-\frac{3}{2}$, and
$A_{-}=-\frac{1}{2}$. 

\begin{figure}
\begin{center}\includegraphics[%
  width=2.8in]{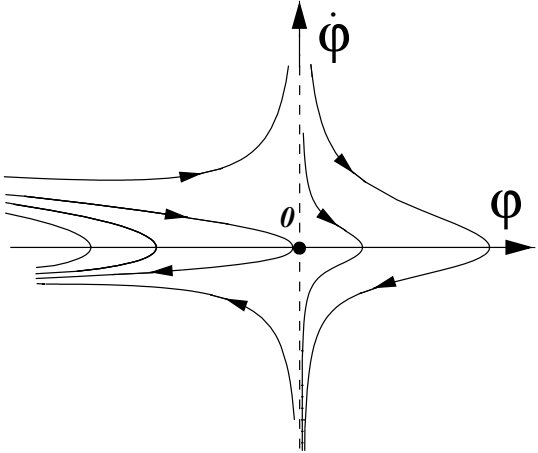}\end{center}

\caption{\label{cap:behavior for minus}The typical behavior of the phase
curves in the neighborhood of the critical line where $K(\varphi)=0$
(here $\dot{\varphi}$ axis ) is plotted for the case when $K_{c}^{\prime}>0$,
$V_{c}^{\prime}=0$, and $V_{c}^{\prime\prime}>0$. }
\end{figure}

Let us now consider such potentials $V(\varphi)$ that $V_{c}^{\prime}=0$.\label{Vstrich 0}
If $K(\varphi)$ is a differentiable function, then, in the case under
consideration, the equation of motion (\ref{eq:fi brifly evolution})
obviously has a fixed-point solution $\varphi(t)\equiv0$ but this
solution is not necessarily the unique one. When $V_{c}^{\prime}=0$
and $K_{c}^{\prime}>0$, then, as follows from the condition (\ref{eq:X-Condition}),
the only value $\dot{\varphi}$, where a phase curve could have coinciding
points with the critical line, is $\dot{\varphi}=0$. From the analysis
made in Sec.~\ref{sub:Tansition-at-X=3D0}, we have already learned
that the transition is impossible in this case. Nevertheless it is
worth to showing explicitly how the phase curves look at this case.
Taking into consideration only the leading order in the numerator
and denominator of Eq.~ (\ref{eq:Phase curve eq.}) and assuming
that $V_{c}^{\prime\prime}\neq0$, we obtain\begin{equation}
\frac{d\dot{\varphi}}{d\varphi}\simeq-\frac{V_{c}^{\prime\prime}}{2\dot{\varphi}K_{c}^{\prime}}.\end{equation}
 The solution of this equation, going through the point $(0,0)$ on
the phase plot, is\begin{equation}
\varphi_{s}=-\dot{\varphi}^{2}\left(\frac{K_{c}^{\prime}}{V_{c}^{\prime\prime}}\right).\label{eq:o-o Trajectory}\end{equation}
In Fig.~\ref{cap:behavior for minus} we have plotted the phase curves
obtained numerically for a toy model with the Lagrangian density $p=\frac{1}{2}\varphi\dot{\varphi}^{2}-\frac{1}{2}\left(\varphi^{2}+2\right)$.
As one can see from Fig.~\ref{cap:behavior for minus}, the parabolalike
phase curve $\varphi_{s}$, given by the formula (\ref{eq:o-o Trajectory}),
is the separatrix going through the fixed-point solution $\varphi(t)\equiv0$.
Moreover, this figure confirms that there are no phase curves intersecting
the critical line by finite $\dot{\varphi}$. \\


\subsection{General differentiable $K(\varphi)$}

The models we are going to discuss below belong to the more general
class of models for which the function $K(\varphi)$ has zero of an
odd order $2n+1$ (where $n\geq0$) at $\varphi=0$. For the linear
in $X$ Lagrangians, we have $\varepsilon_{,XX}\equiv0$; therefore,
if $n>0$ then $K_{c}^{\prime}=0$ and $\varepsilon_{,\varphi X}^{c}=0$.
That is why the general analysis made in the Sec.~ \ref{sub:Transition-at-points}
does not work for this case. Therefore it is interesting to investigate
on this simple example whether the desired transition could be possible
for models not covered by our former analysis. If $K(\varphi)$ is
a sufficient many times differentiable function, then for $\left|\varphi\right|\ll1$
we have\begin{equation}
K(\varphi)\simeq\frac{K_{c}^{\left(2n+1\right)}}{\left(2n+1\right)!}\varphi^{2n+1},\label{eq:K assymptotic}\end{equation}
where $K_{c}^{\left(2n+1\right)}$ is the $(2n+1)$th $\varphi$ derivative
of $K$ at $\varphi=0$. If there is a phase curve, crossing the critical
line at a finite nonvanishing $\dot{\varphi}_{c}$, then integrating
both sides of the equation of motion (\ref{eq:fi brifly evolution})
we obtain\begin{equation}
\dot{\varphi}_{c}-\dot{\varphi}_{in}=-3\int_{\varphi_{in}}^{0}H(\varphi)d\varphi-\int_{\varphi_{in}}^{0}\frac{\varepsilon_{,\varphi}}{\dot{\varphi}K}d\varphi.\label{eq:integral}\end{equation}
Here $(\varphi_{in},\dot{\varphi}_{in})$ is a point on the phase
curve in the neighborhood of the critical line. The first integral
on the right-hand side of Eq.~(\ref{eq:integral}) is always finite,
whereas, as follows from the relation (\ref{eq:K assymptotic}), the
second integral is definitely divergent, if $\varepsilon_{,\varphi}^{c}\neq0$.
This divergence contradicts to our initial assumption: $\dot{\varphi}_{c}$
- finite. Therefore we again obtain the condition (\ref{eq:X-Condition}),
which restricts the possible intersection points on the critical line
in the sense that in the other points, where the condition does not
hold, not only the second derivative $\ddot{\varphi}$ does not exist,
but there are no solutions $\varphi(t)$ at all. Moreover, it is clear
that the condition (\ref{eq:equation on phi dot}) is not enough for
the existence of the solutions intersecting the critical line. Thus,
if the order of $V^{\prime}(\varphi)$ exceeds the order on $\varphi$
of $K^{\prime}(\varphi)$ for $\left|\varphi\right|\ll1$, then one
can neglect $V^{\prime}(\varphi)$ and the integral (\ref{eq:integral})
has the logarithmic divergence (note that we do not consider the points
$\dot{\varphi}_{c}=0$ because as we already know the transition does
not occur via these points). When the order of $V^{\prime}(\varphi)$
is lower than $K^{\prime}(\varphi)$ (and therefore lower than the
order of $K(\varphi)$), we can neglect $\dot{\varphi}^{2}K^{\prime}(\varphi)$
and the integral (\ref{eq:integral}) has a power-low divergence.
Finally, if the functions $K^{\prime}(\varphi)$ and $V^{\prime}(\varphi)$
have the same order on $\varphi$ for $\left|\varphi\right|\ll1$
and are of opposite signs in a sufficient small neighborhood of $\varphi=0$,
then one can find an appropriate finite value $\dot{\varphi}_{c}^{2}\neq0$
for which the divergence on the right-hand side of Eq.~ (\ref{eq:integral})
is canceled. One would expect that at this point the phase curves
intersect the {}``critical'' line and the dark energy changes the
sign of $\left(w+1\right)$. Below, we give the direct calculation
of these $\dot{\varphi}_{c}$ and the phase curves in a neighborhood
of them. Suppose that the order of the functions $\left(V(\varphi)-V_{c}\right)$
and $K(\varphi)$ is $(2n+1)$ and there exist their derivatives of
the order $(2n+2)$. Then for the $\varphi$ derivative of the energy
density we have in the neighborhood of the supposed intersection point
$(0,\dot{\varphi}_{c})$:\begin{align}
\varepsilon_{,\varphi} & \simeq\frac{1}{2}\frac{\varphi^{2n}}{\left(2n\right)!}\left[\left(\dot{\varphi}_{c}^{2}K_{c}^{(2n+1)}+V_{c}^{(2n+1)}\right)+2\dot{\varphi}_{c}K_{c}^{(2n+1)}\delta\dot{\varphi}\right.\nonumber \\
 & +\frac{\varphi}{\left(2n+1\right)}\left.\left(\dot{\varphi}_{c}^{2}K_{c}^{(2n+2)}+V_{c}^{(2n+2)}\right)\right],\label{eq:Assymptotic Energy density}\end{align}
whereas the denominator $\dot{\varphi}K(\varphi)$ of the second integral
on the right-hand side of Eq.~(\ref{eq:integral}) has the order
$(2n+1)$ on $\varphi$. The only possibility to get rid of the divergence
in the integral under consideration is to assume that the first term
in the brackets in the asymptotic (\ref{eq:Assymptotic Energy density})
for $\varepsilon_{,\varphi}$ is zero. Therefore the possible crossing
points are given by \begin{equation}
\dot{\varphi}_{c}=u_{\pm}=\pm\sqrt{-\frac{V_{c}^{(2n+1)}}{K_{c}^{(2n+1)}}.}\end{equation}
Taking into account only the leading order on $\varphi$ and $\delta\dot{\varphi}$
in the denominator and the numerator of Eq.~(\ref{eq:Phase curve eq.}),
we obtain differential equation for the phase curves in the neighborhoods
of the intersection points $(0,u_{\pm})$:

\begin{equation}
\frac{d\delta\dot{\varphi}}{d\varphi}=A_{\pm}-\left(2n+1\right)\frac{\delta\dot{\varphi}}{\varphi},\label{eq:linear on  X Lagra. phase curves general}\end{equation}
where \begin{equation}
A_{\pm}=-3H_{c}+\frac{K_{c}^{(2n+2)}V_{c}^{\left(2n+1\right)}-K_{c}^{(2n+1)}V_{c}^{\left(2n+2\right)}}{2\left[K_{c}^{(2n+1)}\right]^{2}u_{\pm}}.\label{eq:general A}\end{equation}
 The solutions of this equation are given by the formula\begin{equation}
\left(\delta\dot{\varphi}-\frac{A_{\pm}}{2n+2}\varphi\right)\varphi^{2n+1}=const,\end{equation}
which is a generalization of formula (\ref{eq:singular phase curves}).
Similarly to the case $n=0$ ($K_{c}^{\prime}>0$) the solutions,
on which the transition occurs, have the measure zero in the phase
curves set. Therefore we infer that the dynamical transition from
the phantom states with $w\leq-1$ to the usual with $w\geq-1$ (or
vice versa) is impossible. 

Now we would like to mention the models, for which $V^{\prime}(\varphi)$
\emph{}is one order higher on $\varphi$ than $K^{\prime}(\varphi)$
for small $\varphi$. From the asymptotic expression for $\varepsilon_{,\varphi}$
(\ref{eq:Assymptotic Energy density}) and the relation, giving the
possible values of $\dot{\varphi}_{c}$ (\ref{eq:integral}), we see
that the only point on the critical line which could be reached in
a finite time is $\dot{\varphi}_{c}=0$. Therefore, as we have seen
in Sec.~\ref{sub:Tansition-at-X=3D0}, the transition is impossible.
The phase curve going trough the fixed-point solution $\varphi(t)\equiv0$
is a parabola given by the generalization of Eq.~(\ref{eq:o-o Trajectory})
:\begin{equation}
\varphi\simeq-\dot{\varphi}^{2}\left[\frac{K_{c}^{\left(2n+1\right)}}{V_{c}^{\left(2n+2\right)}}\right].\end{equation}
If $V^{\prime}(\varphi)$ is more than one order higher on $\varphi$
than $K^{\prime}(\varphi)$, then as we have already mentioned $\dot{\varphi}_{c}=0$
and the transition is impossible as well.

\subsection{Pole-like $K(\varphi)$}

In this subsection, we briefly consider the case when the function
$K(\varphi)$ has a pole of an odd order, so $K\sim\varphi^{-2n-1}$,
where $n>0$, for $\left|\varphi\right|\ll1$. This kind of functions
$K(\varphi)$ is often discussed in the literature in connection with
the $k-$essence models (see \cite{k-Essence}). Let us keep the same
notation as in subsection \ref{sub:Differentiable-function-}. The
potential $V(\varphi)$ can not have a pole at the point $\varphi=0$,
because, if it were the case, either the energy density $\varepsilon$
or the pressure $p$ would be infinite on the critical line. In order
to obtain finite values of the energy density $\varepsilon$ and pressure
$p$, it is necessary to assume that the system intersects the critical
line at $\dot{\varphi}=0$. But, as we have already seen in Sec.~
\ref{sub:Tansition-at-X=3D0}, the dark energy cannot change the the
sign of $\left(w+1\right)$ at the points $\dot{\varphi}=0$. \\

Thus, we have shown that in the particular case of the theories described
by the linear in $X$ Lagrangians $p(\varphi,X)=K(\varphi)X-V(\varphi),$
which are differentiable in the neighborhood of $\Psi_{c}^{+}$ ($K(\varphi)$
and $V(\varphi)$ differentiable but not necessary linearizable) the
results, obtained for linearizable functions $\varepsilon_{,X}$,
$\varepsilon_{,\varphi}$, and $p_{,X}$, hold as well. This gives
rise to hope that the same statement is true for the general nonlinear
in $X$ Lagrangians as well. Especially we have proven that, if the
construction of the linear in $X$ Lagrangian allows the transition,
then the transitions always realize on a pair of the phase curves.
One phase curve corresponds to the transition from $w>-1$ to $w<-1$
while another one realizes the inverse transition. This pair of phase
curves obviously has the measure zero in the set of trajectories of
the system. Therefore we infer that the considered transition is physically
implausible in this case.

\section{Scalar dark energy in open and closed universes in the presence of
other forms of matter }

In the previous sections, we have seen that the desirable transition
from $w>-1$ to $w<-1$ is either impossible or dynamically unstable
in the case when the scalar dark energy is a dominating source of
gravity in the flat Friedmann universe. Let us now investigate whether
this statement is true in the presence of other forms of matter and
in the cases when the Friedmann universe has open and closed topology.

Following Ref.~\cite{Perturbations}, the effective sound speed $c_{s}$
is given by the same Eq.~(\ref{eq:saund speed}) for the flat, open,
and closed universes. Therefore, if the dark energy is the dominating
source of gravitation (in particular this means that the energy density
of the dark energy $\varepsilon\neq0$ ), then the analysis made in
Sec.~\ref{sub:Transition-at-C^2=3D0} is applicable to open and closed
universes as well as to the flat universe.

\begin{figure}[h]
\begin{center}\includegraphics[%
  bb=0bp 0bp 323bp 213bp,
  width=3in]{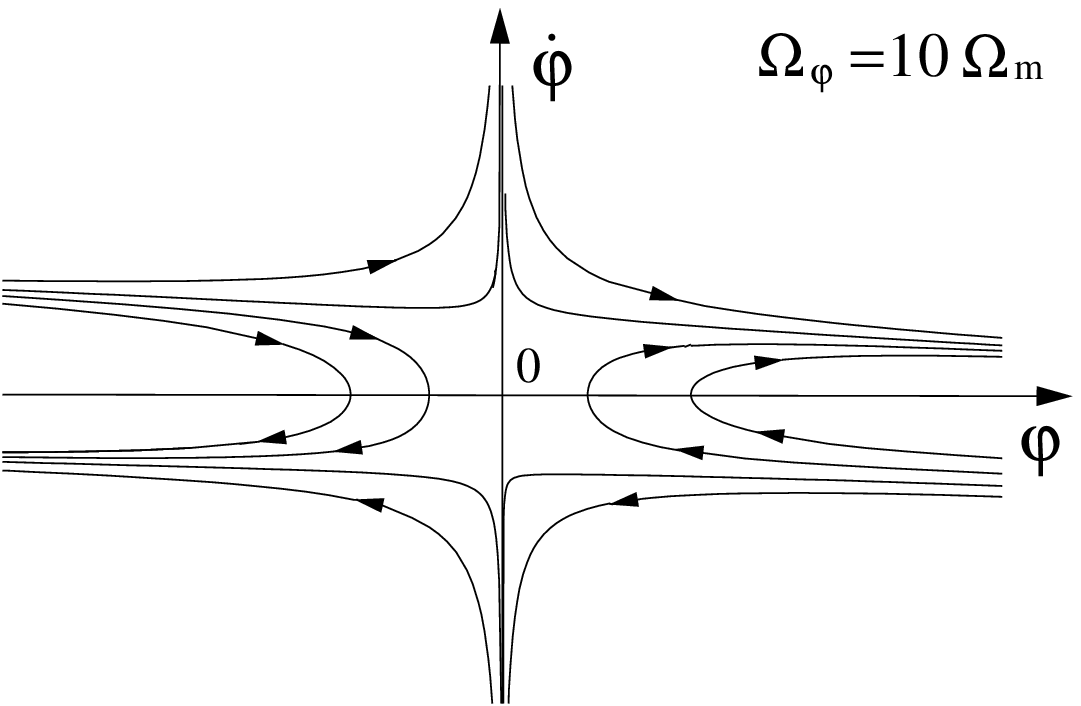}\end{center}

\begin{center}\includegraphics[%
  bb=0bp 0bp 323bp 213bp,
  width=3in]{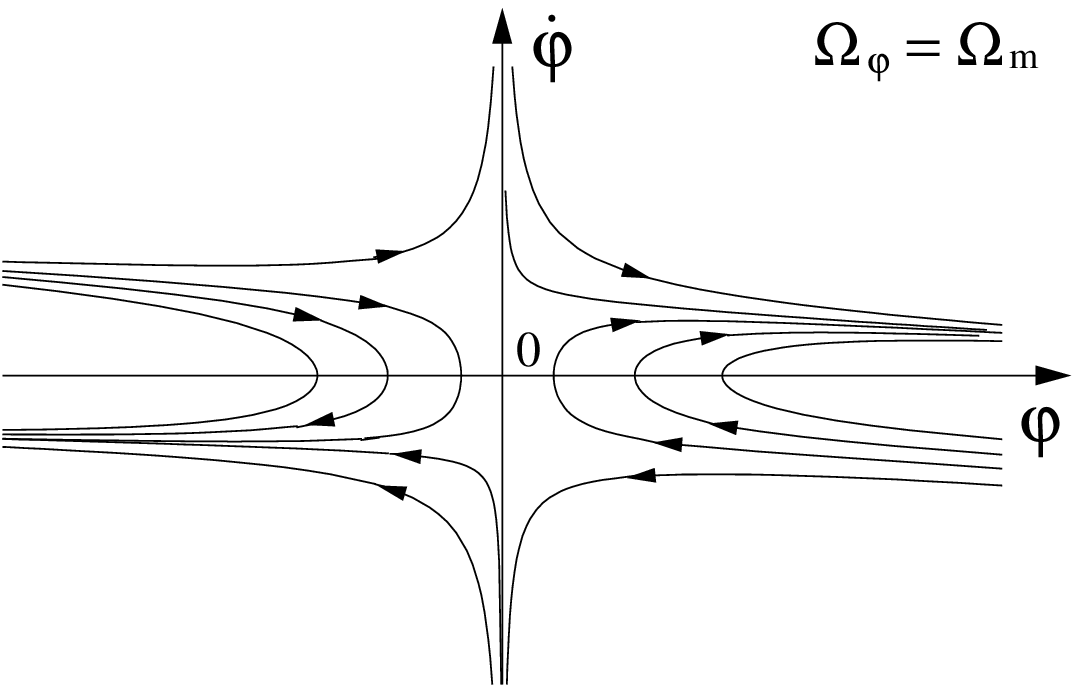}\end{center}

\begin{center}\includegraphics[%
  bb=0bp 0bp 323bp 213bp,
  width=3in]{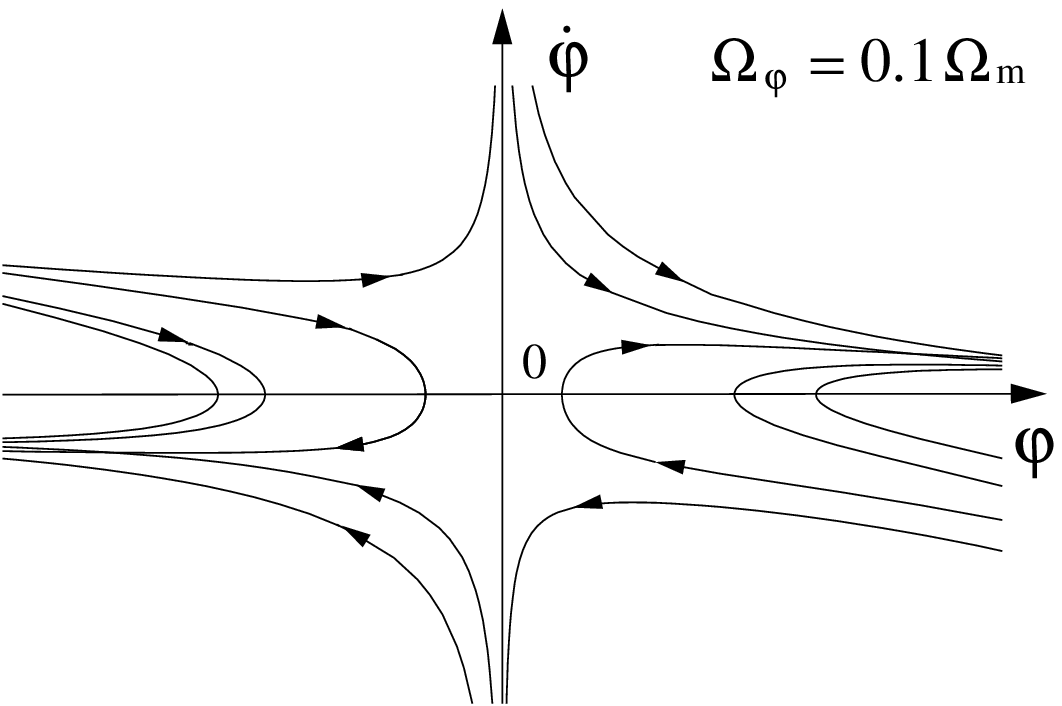}\end{center}

\caption{\label{cap:Trajectories-with Matter}Numerically obtained trajectories
of the dark energy described by a Lagrangian linear in $X$ are plotted
for the cases $\Omega_{\varphi}=10\Omega_{m}$, $\Omega_{\varphi}=\Omega_{m}$,
and $\Omega_{\varphi}=0.1\Omega_{m}$.}
\end{figure}

If the dark energy under consideration interacts with ordinary forms
of matter only through indirect gravitational-strength couplings,
then the equation of motion (\ref{eq:Eq. of moution}) can be written
in the following form: \begin{equation}
\ddot{\varphi}\varepsilon_{,X}+3\dot{\varphi}Hp_{,X}+\varepsilon_{,\varphi}=0,\label{eq:H-eq of motion}\end{equation}
 where merely the Hubble parameter depends on the spatial curvature
and other forms of matter. This dependence is given by the Friedmann
equation:\begin{equation}
H^{2}+\frac{k}{a^{2}}=\frac{1}{3}\left(\varepsilon+\sum\varepsilon_{i}\right),\end{equation}
where $\sum\varepsilon_{i}$ is the total energy density. It is obvious
that the points on the plot $(\varphi,\dot{\varphi})$ considered
in the most of this paper do not define the whole dynamics of the
system anymore and therefore do not define the states of the whole
system. The analysis made in Secs.~\ref{sub:Tansition-at-X=3D0},
\ref{sub:Transition-at-points}, and \ref{Examples} leans only on
the behavior of the scalar filed $\varphi$ and its first $t$ derivative
$\dot{\varphi}$ in the neighborhoods of their selected values, namely,
such as where some of the conditions $p_{,X}=0$, $\dot{\varphi}=0$,
or $\varepsilon_{,X}=0$ etc. hold. For these conditions, the contributions
into the equation of motion (\ref{eq:H-eq of motion}) coming from
the other forms of matter and spatial curvature would be of a higher
order and therefore are not important for the local behavior of $\varphi$
and the problem as a whole. In fact, the value of the Hubble parameter
did not change the qualitative futures of the phase curves considered
in Secs.~\ref{sub:Tansition-at-X=3D0},\ref{sub:Transition-at-points},
and \ref{Examples}. To illustrate this statement, we plot the trajectories
of the system $p=\frac{1}{2}\varphi\dot{\varphi}^{2}-\frac{1}{2}\left(\left(\varphi-1\right)^{2}+\frac{1}{3}\right)$
(it is the same system that we considered in previous subsection)
in presence of dust matter for various values of the initial energy
densities of the dust (see Fig.~\ref{cap:Trajectories-with Matter}).
The only thing that is important is that $H\neq0$. The universe should
not change the expansion to the collapse and the plot of the scale
factor $a(t)$ should not have a cusp directly at the time of the
transition. Thus, we infer that the most of our analysis is applicable
to a more general physical situation of a Friedmann universe filled
with various kinds of usual matter, which interact with the dark energy
only through indirect gravitational-strength couplings. Moreover,
if the interaction between the dark energy field $\varphi$ and other
fields does not include coupling to the derivatives $\nabla_{\mu}\varphi$,
then the obtained result holds as well. %

\section{Conclusions and Discussion}

In this paper we have found that the transitions from $w>-1$ to $w<-1$
(or vice versa) of the dark energy described by a general scalar-field
Lagrangian $p(\varphi,\nabla_{\mu}\varphi)$ are either unstable with
respect to the cosmological perturbations or realized on the trajectories
of the measure zero. If the dark energy dominates in the universe,
this result is still robust in the presence of other energy components
interacting with the dark energy through nonkinetic couplings. In
particular, we have shown that, under this assumption about interaction,
the dark energy described by Lagrangians linear in $\left(\nabla_{\mu}\varphi\right)^{2}$
cannot yield such transitions even if it is a subdominant source of
gravitation.

Let us now discuss the consequences of these results. If further observations
confirm the evolution of the dark energy dominating in the universe,
from $w\geq-1$ in the close past to $w<-1$ to date, then it is impossible
to explain this phenomenon by the classical dynamics given by an effective
scalar-field Lagrangian $p(\varphi,\nabla_{\mu}\varphi)$. In fact,
the models which allow such transitions have been already proposed
(see e.g.,~\cite{Triad,Xinmin Zhang,Shtanov} and other models from
the Ref.~\cite{Review}) but they incorporate more complicated physics
then the classical dynamics of a one scalar field. 

If observations reveal that $w<-1$ now and if we disregard the possibility
of the transitions, then the energy density of the dark energy should
grow rapidly during the expansion of the universe and therefore the
coincidence problem becomes even more difficult. Thus, from this point
of view the transitions considered in this paper would be rather desirable
for the history of the universe. As we have shown, to explain the
transition under the minimal assumptions of the nonkinetic interaction
of dark energy and other matter one should suppose that the dark energy
was subdominating and described by a nonlinear in $\left(\nabla_{\mu}\varphi\right)^{2}$
Lagrangian. Thus, some nonlinear (or probably quantum) physics must
be invoked to explain the value $w<-1$ in models with one scalar
field.

The second application of our analysis is the problem of the cosmological
singularity. To obtain a bounce instead of collapse, the scalar field
$\varphi$ must change its equation of state to the phantom one before
the bounce and should dominate in the universe at the moment of transition.
Otherwise, if the scalar field was subdominant then it is still subdominant
after the transition as well, because its energy density decreases
during the collapse, while the other nonphantom forms of matter increase
their energy densities. The disappearing energy density of $\varphi$
does not affect the gravitational dynamics and therefore does not
lead to the bounce. On the other hand, as we have already proved,
a dominant scalar field $\varphi$ described by the action without
kinetic couplings and higher derivatives cannot smoothly evolve to
the phantom with $w<-1$. Therefore we infer that a smooth bounce
of the nonclosed Friedmann universe cannot be realized in this framework.

\begin{acknowledgments}
It is a pleasure to thank Slava Mukhanov and Serge Winitzki for useful
and stimulating discussions and for their helpful comments on the
draft of this paper. 
\end{acknowledgments}

\end{document}